\documentclass[preprint,amsmath,amssymb,prl,lettersizepaper,tightenlines,superscriptaddress]{revtex4-1}
\usepackage{amsmath,amssymb}
\usepackage{xcolor}
\usepackage{graphicx}
\usepackage{subfigure}
\usepackage{subfig}
\usepackage{dcolumn}
\usepackage{bm}
\usepackage{float} 
\usepackage{amsmath}
\usepackage{amsmath, amsthm, amsfonts}

\usepackage{dsfont}
\usepackage{amsmath}
\usepackage[utf8]{inputenc}
\usepackage{calrsfs}
\usepackage[mathcal]{eucal}
\usepackage{multirow}
\DeclareGraphicsExtensions{.bmp}

\newcommand{\be}{\begin{equation}}
\newcommand{\ee}{\end{equation}}

\newcommand{\bit}{\begin{itemize}}
	\newcommand{\eit}{\end{itemize}}

\graphicspath{{Figures//}}
\usepackage{graphicx}
\begin{document}
	\title{\large{A solid-state emitter embedded in a microcavity under intense excitation: a variational master equation approach}}
	
	\author{Oscar J. Gómez-Sánchez}
	\affiliation{Grupo de F\'isica Te\'orica y Computacional, Escuela de F\'isica, Universidad Pedag\'ogica y Tecnol\'ogica de Colombia (UPTC), Tunja 150003, Boyac\'a, Colombia.}
	\affiliation{Department of Electrophysics, National Chiao Tung University (NCTU), Hsinchu 30050, Taiwan}
	\author{Hanz Y. Ram\'irez}
	\email{hanz.ramirez@uptc.edu.co}
	\affiliation{Grupo de F\'isica Te\'orica y Computacional, Escuela de F\'isica, Universidad Pedag\'ogica y Tecnol\'ogica de Colombia (UPTC), Tunja 150003, Boyac\'a, Colombia.}

	\date{\today}
	
	\begin{abstract}
		In this work, dissipative effects from a phonon bath on the resonance fluorescence of a solid-state two level system embedded in a high quality semiconductor microcavity and driven by an intense laser, are investigated.
		Within the density operator formalism, we derive a variational master equation  valid for broader ranges of temperatures, pumping rates, and radiation-matter couplings, than previous studies. From the obtained master equation, fluorescence spectra for various thermal and exciting conditions are numerically calculated, and compared to those computed from weak coupling and polaronic master equations, respectively. Our results evidence the breakdown of those rougher approaches under increased temperature and strong pumping.  
	\end{abstract}
	\maketitle
\section{I. INTRODUCTION}

Solid-state emitters embedded in microcavities have become a new paradigm in cavity quantum electrodynamics (cavity-QED) \cite{murch2017cavity,yoshihara2017superconducting,forn2017ultrastrong}.
Recent developments in fabrication of semiconductor cavities serve a number of research fields, including quantum information processing, photonic circuits and quantum optics \cite{heo2017scheme,chiorescu2004coherent,Coles,Stockklauser,Dufferwiel}. Regarding the later, high quality cavities have been crucial for boosting the efficiency of single photon generators \cite{song2017ingaasp,ding2016demand,Unsleber}.

For instance, some late experimental studies have focused on the resonant fluorescence of InGaAs quantum dots (QDs) grown inside of microcolumns, which have provided a clear demonstration of induced excitation \cite{Wang,ulrich2011dephasing,Hargart}. Thus, in systems with non-resonant laser-cavity coupling, the cavity mode is indirectly excited by the emission of photons from an artificial atom coupled to the acoustic phonon environment (phonon assisted cavity feeding) \cite{ates2009post,calic2011phonon}. The inverse effect of non-resonant coupling, where the quantum emitter is excited by photons emitted from the cavity, has also been observed \cite{majumdar2011phonon}.

D. McCutcheon {\it et al.} developed some years ago a variational master equation to describe the dynamics of a cavityless two-level system interacting with a boson environment, which was applied in the study of Rabi's rotations of a quantum dot \cite{mccutcheon2011general}. They found that the technique of variational master equation captures effects generally considered non-perturbative, such as multiphoton processes and renormalization of the Rabi frequency induced by the phonon bath. By comparing their population dynamics results with path integral numerical calculations, the reliability of the variational approach in accounting for those non-perturbative effects in regimes in which the weak and polaronic models was verified. 

Nevertheless, state of the art experiments use optical resonators embedding the emitter, because of the associated improvement in collection rates and photon purity \cite{PRLpolariton,stanford2016,ding2016demand,Seyfferle,hefei2018,experiment2018}. Thus, our purpose is to investigate the fluorescence spectrum  of a solid-state qubit-cavity system under pumping and thermal conditions beyond the scope of previously studied formulations like the weak coupling and polaronic approaches. To do that, we derive a variational master equation, which allows for numerical simulations of resonance fluorescence spectra within a wider range of excitation rates, emitter-cavity couplings, and temperatures. Such a master equation might also contribute to the promising research thread on double-dot-cavity systems, regarding phonon dissipation in tunnel-coupled emitters \cite{petta2014,petta2015,unal2016,proceeding2017,dipolaritons2018}.

Although this kind of systems have been addressed by means of numerical approaches, which adequately implemented, may render a solution as close as desired to the exact one (e.g. quasi-adiabatic propagator path-integral or real-time path integral techniques) \cite{JoCP2001,PRL2007,mccutcheon2011general,PRB2011,JoPCM2017}; those techniques are highly demanding from the computational point of view and do not yield the physical insight provided by a master equation.

 This paper is organized as follows:  In the next section we present the model Hamiltonian and its modification under an adequate unitary transformation. In section  III,  the free energy of the system is minimized to determine the variational parameters and in section IV, the corresponding variational master equation is derived. Finally, in section V we obtain and discuss numerical simulations of fluorescence spectra of a semiconductor QD coupled to a cavity mode, and end by drawing overall conclusions in section VI.

\section{II. THEORY}
	
The system under study is a solid-state two level system (which we will refer to as ``quantum dot'' although it could either be a vacancy in a 3D crystal, a localized defect in a low dimensional structure, a nanocrystal, or any other suitable artificial atom), embedded in a QED cavity \cite{microcavitieKreinberg}.

\begin{figure}
	\centering
	\includegraphics[width=0.5 \linewidth]{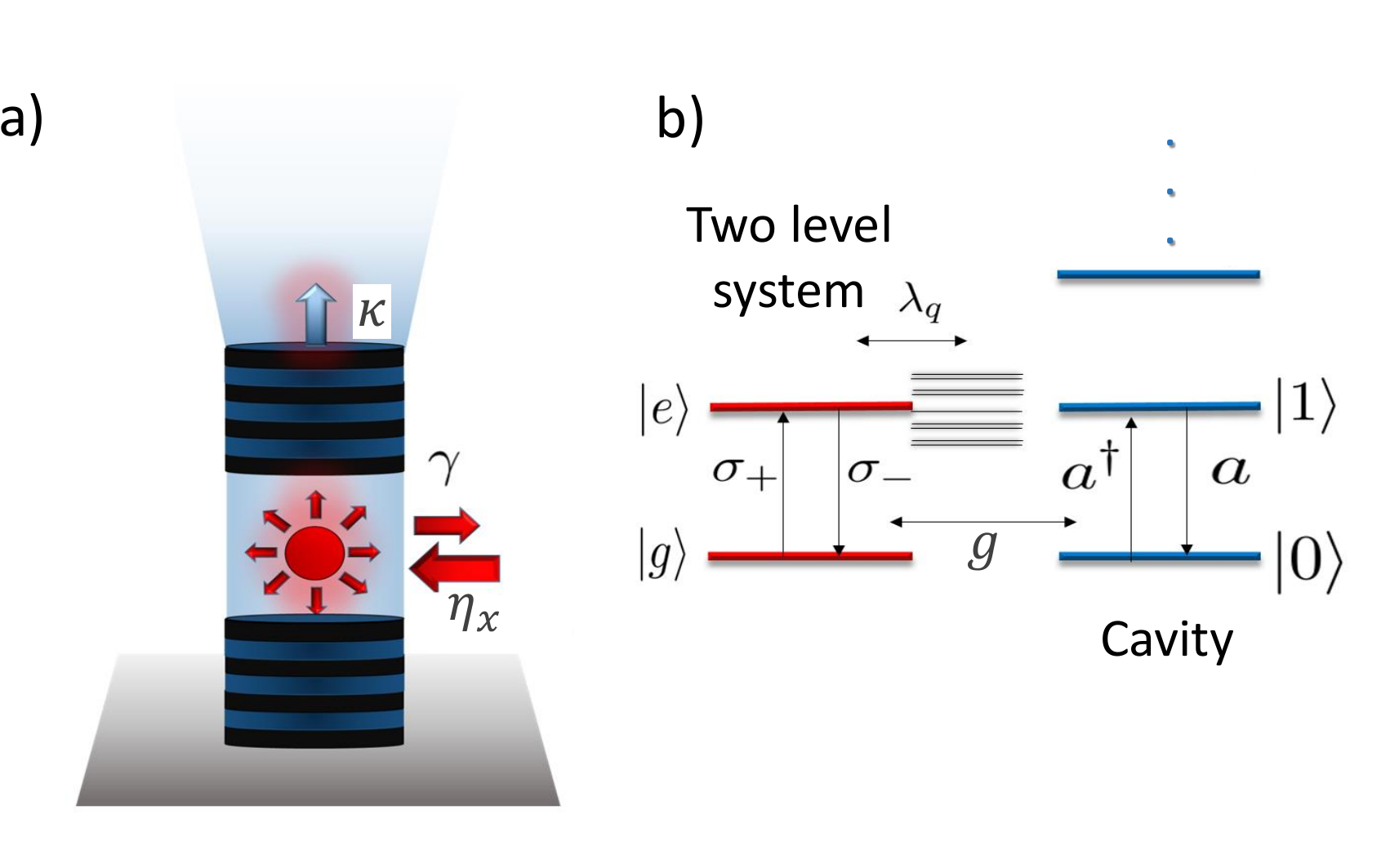}
	\caption{ a) Schematics of a quantum emitter (pure radiative linewidth $\gamma$), embedded in a micropillar cavity (loss rate $\kappa$), and driven by a laterally applied CW laser (pumping rate $\eta_x$). 		b) Emitter energy levels (ground $| g \rangle$ and excited $| e \rangle $), and its interactions with the phonon reservoir ($\lambda_q$) and the cavity ($ g $). }
	\label{fig:modelo_a_b}
\end{figure}

Carriers confined in the QD interact with a continuum of states in the sample of which it is part, via  acoustic phonons. This interaction causes an incoherent pumping of the two level system. Moreover, because the artificial atom mainly interacts with a cavity single mode, the phonon environment  produces some decoherence effects in the atom-cavity arrangement. The system is assumed driven by a continuous wave (CW) laser, as shown in the figure \ref{fig:modelo_a_b}a), while the corresponding energy levels and interactions are depicted in figure \ref{fig:modelo_a_b}b).

Working in a rotating frame whose frequency matches that of the exciting laser $\omega_L$ \cite{dieter,RoyHam}, the considered Hamiltonian reads ($\hbar=1$)

\begin{equation}
\begin{split}
\hat{H}&=  \Delta_{XL}\hat{\sigma}^+\hat{\sigma}^- 
+  \Delta_{CL}\hat{a}^{\dagger}\hat{a}\\
& +  \eta_x(\hat{\sigma}^+ + \hat{\sigma}^-) +   g(\hat{\sigma}^+ \hat{a} + \hat{a}^\dagger \hat{\sigma}^-)\\
& + \hat{\sigma}^+ \hat{\sigma}^- \sum_{q}  \lambda_{q} (\hat{b}_{q} + \hat{b}_{q}^\dagger) 
+\sum_{q}  \omega_{q} \hat{b}_{q}^\dagger \hat{b}_{q} \hspace{1ex} ,
\label{H1}
\end{split}
\end{equation}

where $\omega_{q}$ is the frequency of a phonon with momentum $q$, while $b_{q}$  ($b_{q}^\dagger$) and $\lambda_{q}$ are correspondingly the boson annihilation (creation) operator and intensity of the carrier-phonon coupling. The detuning respect to the pumping laser of the two level transition frequency ($\omega_X$) and that of the cavity mode ($\omega_C$), are respectively $\Delta_{C L}$ and $\Delta_{X L}$. The annihilation  (creation) operator of photons at the cavity frequency is $\hat{a}$ ($\hat{a}^{\dagger}$), while the QD dipole operators are $\sigma^-$ and $\sigma^+$. $g$ is the radiation-matter coupling constant, and the pumping rate $\eta_x$ is the half of the Rabi frequency associated to the driving laser power.

 Let us consider a generalization of the polaron transformation that displaces the phonon bath oscillators,  by an amount that is determined by a set of variational parameters $\{f_q\}$ \cite{mccutcheon2011general}. Such a variational transformation can be written as
 
\begin{equation}
\hat{H}'=e^{\hat{S}} \hat{H} e^{-\hat{S}} \hspace{1ex} ,
\end{equation}

where

\begin{equation}
	\hat{S}=\hat{\sigma}^+\hat{\sigma}^-\sum_q \nu_q (\hat{b}_q^{\dagger}-\hat{b}_q) \hspace{1ex} ,
\end{equation}

where $\nu_q = \frac{f_q}{\omega_q}$.

The transformed Hamiltonian becomes $\hat{H}'_S + \hat{H}'_I + \hat{H}'_B$, with

\begin{align}
&\hat{H}'_S=  \Delta_{R}\hat{\sigma}^+\hat{\sigma}^- 
+  \Delta_{CL}\hat{a}^{\dagger}\hat{a}+   \langle \hat{B}\rangle \hat{\zeta}_x \hspace{1ex} ,\\
&\hat{H}'_I=\sum_{i=x,y,z}   \hat{\zeta}_{i} \hat{B}_i \hspace{1ex} , \\
&\hat{H}'_B=\sum_{q}  \omega_q\hat{b}_q^\dagger \hat{b}_q \hspace{1ex} ,
\label{Hw}
\end{align}

where the modified detuning $\Delta_R=\Delta_{XL}+R$, depends on the variational  shift $ R=\sum_q\omega^{-1}_qf_q(f_q-2\lambda_q)$, 
and the thermal average of the bath displacement operator is given by ($\beta=1/k_BT$) 

\begin{equation}
\langle \hat{B}\rangle =\exp\left[-\frac{1}{2}\sum_{q}\frac{f^2_q}{\omega_q^2}\coth (\beta  \omega_q/2)\right] \hspace{1ex} .
\end{equation}

In turn, the system modified operators $\hat{\zeta}_{i}$ are explicitly 

\begin{align}
&\hat{\zeta}_x=\eta_x(\hat{\sigma}^++\hat{\sigma}^-)+g(\hat{\sigma}^+\hat{a}+\hat{\sigma}^-\hat{a}^\dagger) \hspace{1ex} ,\\
&\hat{\zeta}_y=i\eta_x(\hat{\sigma}^+-\hat{\sigma}^-)+ig(\hat{\sigma}^+\hat{a}-\hat{\sigma}^-\hat{a}^\dagger) \hspace{1ex} ,\\
&\hat{\zeta}_{z}=\hat{\sigma}^+\hat{\sigma}^- \hspace{1ex} ,
\end{align}

and the phonon-induced fluctuation operators are defined as


\begin{align}
& \hat{B}_x=\frac{1}{2}(\hat{B}_+ + \hat{B}_- - 2\langle \hat{B}\rangle ) \hspace{1ex} ,\\
& \hat{B}_y=\frac{1}{2i}(\hat{B}_+-\hat{B}_-) \\
&\hat{B}_z=\sum_q(\lambda_q-f_q)(\hat{b}^\dagger_q+\hat{b}_q) \hspace{1ex} ,
\label{B_yB_x V}
\end{align}

in terms of the coherent displacement operators 

\begin{align} 
&\hat{B}_\pm=e^{\pm\sum_{q}\nu_q(\hat{b}_q^\dagger-\hat{b}_q)} \hspace{1ex} .
\end{align}

In the limit of continuous phonon modes, which is convenient and appropriate as long as the lattice parameter is much smaller than the typical size of the sample embedding the emitter, a spectral density $J(\omega)$ must be introduced, so that $\langle B\rangle$ and $R$ correspondingly turn into

\begin{align}
&R=\int_0^\infty d\omega \hspace{1ex} \frac{J(\omega)}{\omega} F(\omega)(F(\omega)-2),\\
&\langle B\rangle =\exp\left[-\frac{1}{2}\int_0^\infty d\omega \hspace{1ex} \frac{J(\omega)^2F(\omega)^2}{\omega^2}\coth(\beta  \omega/2)\right] \hspace{1ex}.
\end{align}


\section{III. Free energy minimization}

 The variational parameters $\{ f_q \}$ must be chosen in such a way that they minimize the free energy associated
 with the transformed Hamiltonian \cite{silbey1984variational,kuzemsky2015variational,nazir2016modelling}. To do that, we use the Feynman–Bogoliubov inequality $A_u\geq A$, according which the free energy of the system ($A$), is at first order bounded by an upper limit given by
 
 \begin{equation}
 \begin{split}
 A_u=-\frac{1}{\beta}\ln\left(\text{Tr}\Big\{e^{-\beta \hat{H}'_{0}}\Big\}\right)+\langle \hat{H}'_{I}\rangle _{\hat{H}'_{0}} \hspace{1ex},
 \label{F_B bound}
 \end{split}
 \end{equation}
 
  where  $ \hat{H}'_{0}= \hat{H}'_{S}+\hat{H}'_{B}$ and  $\langle \hat{H}'_{I}\rangle _{\hat{H}'_{0}}=\text{Tr}\{\hat{H}'_{I}e^{-\beta \hat{H}'_{0}}\}$.	
  
 On the one hand, $\langle \hat{H}'_{I}\rangle _{\hat{H}'_{0}}$ vanishes because in the basis of eigenstates of $\hat{H}'_{0}$, all diagonal terms of $\hat{H}'_{I}$ are zero. On the other hand, since $[\hat{H}'_{B},\hat{H}'_{S}]=0$ and each of those operators act on eigenstates of different subspaces (the dot-cavity and the phonon bath), then $A_u$ can be reduced to
  
  \begin{equation}
  \begin{split}
  A_u=A_B-\frac{1}{\beta}\ln\left(\text{Tr}\Big\{e^{-\beta \hat{H}'_{S}}\Big\}\right)
  \label{A_u} \hspace{1ex},
  \end{split}
  \end{equation}
    
  with $A_B$ the free energy of the phonon bath. 
  Inserting equation (4) into equation (\ref{F_B bound}), the Feynman–Bogoliubov upper bound reads 
  
\begin{eqnarray}
  	A_u =& A_B - \frac{1}{\beta} \ln [ 2e^{-\frac{\beta}{2}((2n-1)\Delta_{CL}+\Delta_R)} \\
  	&\times \left(\cosh\left[\frac{1}{2}\beta\mu_1\right]+\cosh\left[\frac{1}{2}\beta\mu_2\right]\right)]  \hspace{1ex},
  	\label{A_u2}
\end{eqnarray}

in terms of the phonon mean occupation number at temperature T 
($n=\langle\hat{b}^{\dagger}\hat{b}\rangle=\left[e^{\beta \omega}-1\right]^{-1}$), and of the 
quantities 

\begin{equation}
\begin{split}
&\mu_1=\sqrt{\mho_1 +2 \mho_2} \hspace{1ex} , \hspace{1ex} \mu_2=\sqrt{\mho_1 - 2\mho_2 } \hspace{1ex},
\end{split}
\end{equation} 

that in turn depend on

\begin{equation*}
\begin{split}
&\mho_1=\Delta_{CL}^2 + \Delta_R^2 +2 B^2 (g^2 n + 2 \eta_x^2) \hspace{1ex}, \\
&\mho_2=\sqrt{(B^2 g^2 n - \Delta_{CL} \Delta_R)^2 + 	4 B^2 (B^2 g^2 n + \Delta_{CL}^2) \eta_x^2} \hspace{1ex}.
\end{split}
\end{equation*}

Because the free energy of the phonon bath does not depend on $f_q$, i.e. $ A_B $ is unchanged by the interaction with the system, and then it is irrelevant in minimizing  $A_u$. By imposing $\frac{\partial A_u}{\partial f_q} = 0$, we obtain 

{\tiny
	
\begin{widetext}
	\begin{equation}
	f_q\equiv\lambda_q F(\omega_q) = \frac{\lambda_q\left(1-\frac{\frac{\Delta R +\Lambda 2}{\mu_1}\sinh\left(\beta \mu_1/2\right)+\frac{\Delta R -\Lambda 2}{\mu_2}\sinh\left(\beta \mu_2/2\right)}{\cosh\left(\beta\mu_1/2\right)+\cosh\left(\beta \mu_2/2\right)}\right)}{
		1-\frac{\left(\frac{\Delta R +\Lambda 2}{\mu_1}\sinh\left(\beta \mu_1/2\right)+\frac{\Delta R -\Lambda 2}{\mu_2}\sinh\left(\beta \mu_2/2\right)\right)}{\cosh\left(\beta \mu_2/2\right)+\cosh\left(\beta\mu_2/2\right)}+\frac{B^2}{\omega_q}\frac{
			\left( \frac{(ng^2+2\eta_x^2)+\Lambda_1}{\mu_1}\sinh\left(\beta \mu_1/2\right)+\frac{(ng^2+2\eta_x^2)-\Lambda_1}{\mu_2}\sinh\left(\beta \mu_2/2\right) \right)}{\cosh\left(\beta \mu_2/2\right)+\cosh\left(\beta\mu_2/2\right)}\coth\left(\beta\omega_q/2\right)} \hspace{1ex},
	\label{F} 
	\end{equation}
\end{widetext} }

where $\Lambda_1=\frac{B^2g^2n(g^2n+4\eta_x^2)-\Delta_{CL}(g^2n\delta_R-2\Delta_{CL}\eta_x^2)}{\mho_2} $
and $\Lambda_2=\frac{\Delta_{CL}(\Delta_{CL}\Delta_R-B^2g^2n)}{\mho_2}$.

\begin{figure}[H]
	\begin{center}
	\includegraphics[width=0.5\linewidth]{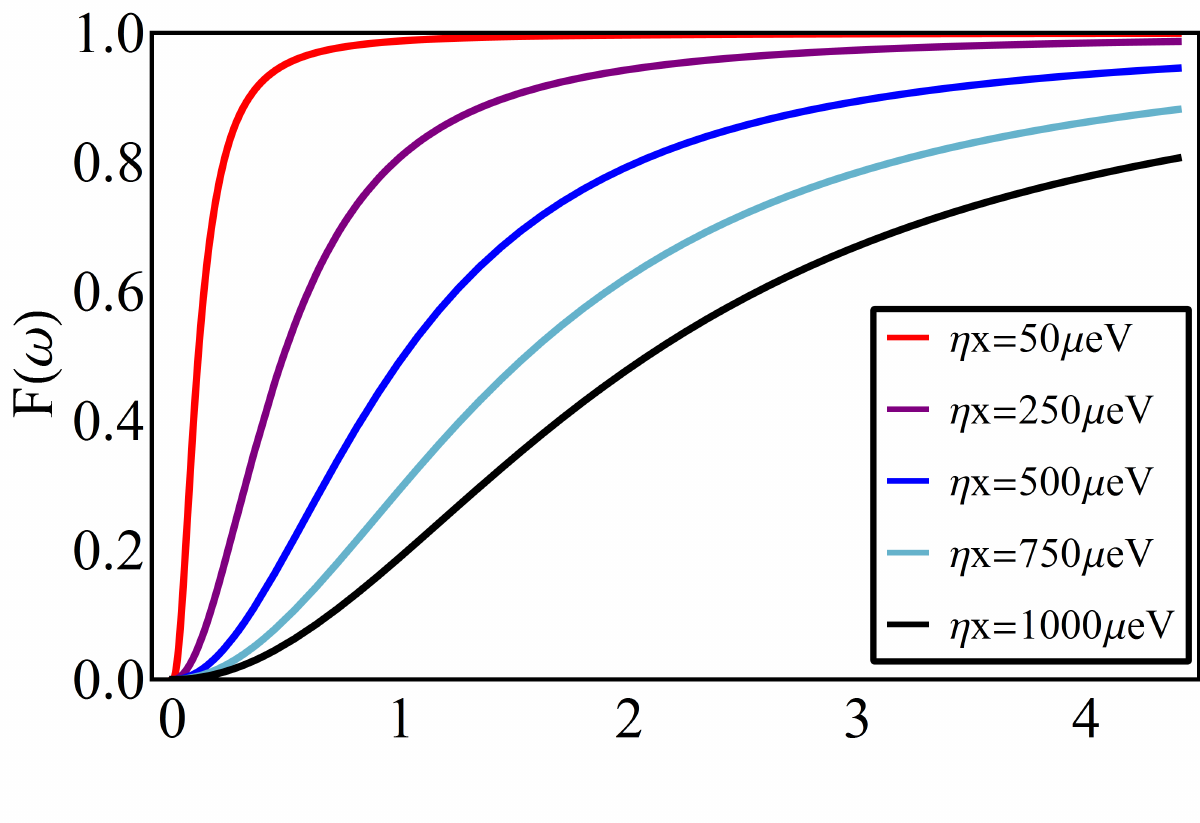}\\
	\includegraphics[width=0.5\linewidth]{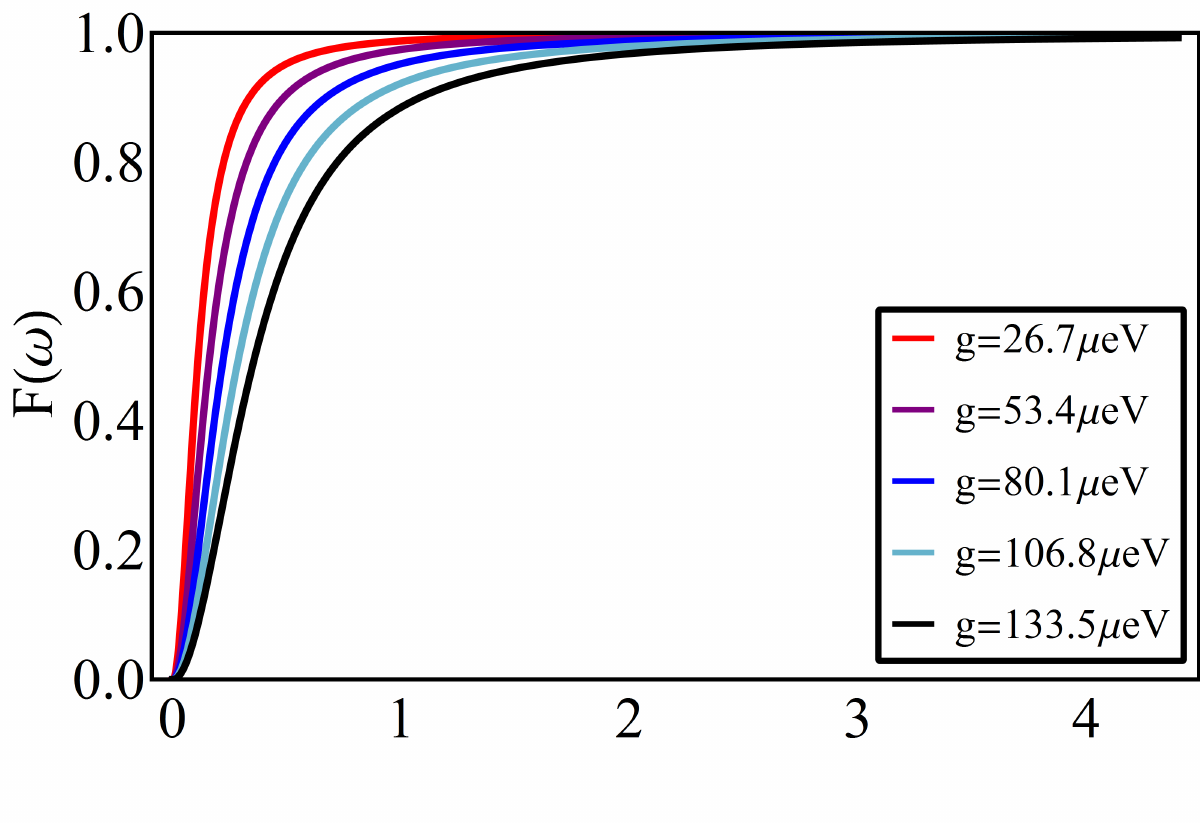}\\
	\includegraphics[width=0.5\linewidth]{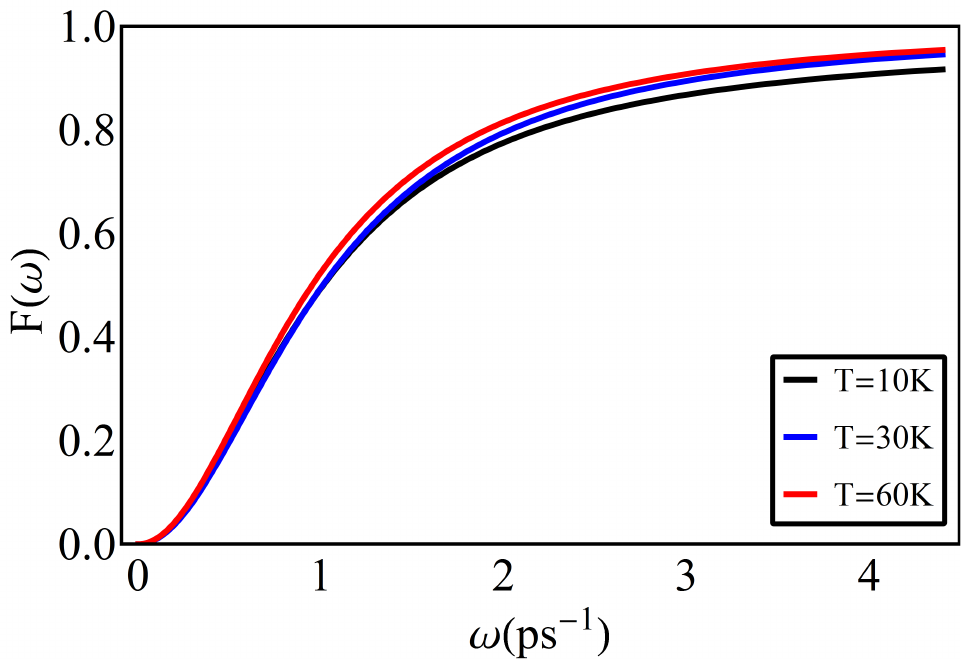}
	\caption{ a) Variational parameter as function of the phonon frequency, at $T=30$ K and $g=26.7$ $\mu$eV for different pumping rates (Upper panel: from bottom to top, the curves correspond to smaller rates), b) at $T=30$ K and $\eta_x=100$ $\mu$eV for different coupling constants (Middle panel: from bottom to top, the curves correspond to smaller couplings), and c) at  $\eta_x=500$ $\mu$eV and  $g=26.7$ $\mu$eV for different temperatures (Lower panel: from bottom to top, the curves correspond to higher temperatures). }
	\label{parametrof}
	\end{center}
\end{figure}

In figure \ref{parametrof} the frequency dependence of the modulating part of the variational parameters for different pumping rates, radiation-matters couplings and temperatures is presented. There can be seen how for wave vectors $q$ whose associated frequencies satisfy $\eta_x/\omega_q\ll 1$, the minimization condition yields $f_q \rightarrow \lambda_q$, recovering the polaronic limit \cite{nazir2016modelling}. Only for these modes, the bath oscillators can fully follow the atom excitation. Otherwise, the mode frequencies are too slow and the corresponding oscillator shifts are dwindled, so that the carrier-phonon coupling at the corresponding momentum range is inhibited.

\section{IV. Master equation}
 
In this section, a variational master equation for the reduced density operator $\hat{\rho}(t)$, of the QD-cavity system, is derived within the second order Born-Markov framework \cite{schlosshauer2007decoherence}. The use of those approximations is justified because even at room temperature, the thermal energy would be much smaller than the typical transition energy of the two level emitter, and the thermalization processes are much faster than the relevant optical dynamics \cite{Royeff}. The validity of convolutionless non-perturbative approaches (regarding the phonon-carrier interaction) for studying strongly coupled dot-cavity systems, has been shown in references \cite{roy2011phonon,RoyHam}. In the case of strong pumping, minimization of the free energy is expected to grab relevant non-Markovian effects.  

We include the emitter radiative recombination and the cavity losses as Liouvillian decay superoperators, which act on the density matrix of the reduced system \cite{Carmichae}. Such operators in the Lindblad form are given by  

 \begin{equation}
\begin{split}
\mathcal{L}(\hat{\rho})=&\frac{\gamma}{2}\left( 2\hat{\sigma}^-\hat{\rho}\hat{\sigma}^+-\hat{\sigma}^+\hat{\sigma}^-\hat{\rho}-\hat{\rho}\hat{\sigma}^+\hat{\sigma}^-\right)\\
&+\kappa\left( 2\hat{a}\hat{\rho}\hat{a}^{\dagger}-\hat{a}^{\dagger}\hat{a}\hat{\rho}-\hat{\rho}\hat{a}^{\dagger}\hat{a}\right) \hspace{1ex}\\
\end{split}
\end{equation}

where $\gamma/2$ is the HWHM radiative linewidth and $\kappa$ is the cavity loss rate for the relevant mode.  

Thus, inserting the transformed Hamiltonian from equations (4), (5) and (6), the variational master equation takes the form
	 
\begin{equation}
\begin{split}
\frac{\partial \hat{\rho}}{\partial t}=& -i[H_s,\hat{\rho}(t)]+\mathcal{L}(\hat{\rho})\\
& -  \int_{0}^{t}d\tau \mathop{\sum_{l=x,y,z}}_{ m=x,y,z} C_{lm}[\hat{\zeta}_m,e^{-iH_s\tau} \hspace{1ex} \hat{\zeta}_l \hspace{1ex} e^{iH_s\tau  }\hat{\rho}(t)\\
&+ \int_{0}^{t}d\tau\mathop{\sum_{l=x,y,z}}_{ m=x,y,z} C_{lm}^*[\hat{\rho}(t)e^{-iH_s\tau} \hspace{1ex} \hat{\zeta}_l \hspace{1ex} e^{iH_s\tau },\hat{\zeta}_m] \hspace{1ex},
\label{simplifiedME}
\end{split}
\end{equation}

where $C_{lm}(\tau)=\langle B_l(\tau)B_m\rangle$ for $l,m=x,y,z$. 

Assuming that the phonon bath is in thermal equilibrium \cite{glauber1963coherent}, the correlation
 functions become
  
	 \begin{align}
	 & C_{yy}(\tau)=\langle B\rangle ^2 \left(\cos\phi(\tau)-1\right) \hspace{1ex}, \nonumber\\
	 & C_{xx}(\tau)=\langle B\rangle ^2 \sin\phi(\tau) \hspace{1ex}, \nonumber \\
	 & C_{zz}(\tau)=\int_0^\infty d\omega J(\omega)[1-F(\omega)]^2  \nonumber\\
	 &\ \ \ \ \ \ \ \ \  \times(\cos\omega\tau\coth( \beta  \omega/2)-i\sin\omega\tau) \hspace{1ex}, \nonumber\\
	  &C_{zy}(\tau)=\langle B\rangle \int_0^\infty d\omega\frac{J(\omega)}{\omega}F(\omega)[1-F(\omega)] \nonumber \\ &\ \ \ \ \ \ \ \ \ \times\left(i\cos\omega\tau+\sin\omega\tau\coth(\beta  \omega/2)\right) \hspace{1ex}, \nonumber \\
	  &C_{yz}(\tau)=-\langle B_z(\tau)B_y(0)\rangle \hspace{1ex}, \nonumber \\
	  \text{and} \nonumber \\
	 & C_{xz}(\tau)=C_{zx}(\tau)=C_{xy}(\tau)=C_{yx}(\tau)=0 \hspace{1ex} ,
	 \end{align}
	 
	 which depend on the spectral density and on the variational parameters. The first two correlations also depend on the function
	 
	 \begin{equation}
	 \phi(\tau)=\int_0^\infty d\omega \frac{J(\omega)}{\omega^2}F(\omega)^2\left(\cos\omega\tau\coth(\beta\omega/2)-i\sin\omega\tau\right) \hspace{1ex} .
	 \end{equation}

On the other hand, the master equation can be written in the Lindblad form 

\begin{equation}
\begin{split}
\frac{\partial\hat{\rho}(t)}{\partial t}
=&-i\left(\left[H_{S}^{ef},\hat{\rho}(t)\right]+D_{ph}(\hat{\rho})
\right)+\mathcal{L}(\hat{\rho})+
\mathcal{L}_{ph}(\hat{\rho}) \hspace{1ex},
\label{VME}
\end{split}
\end{equation}

in terms of the effective Hamiltonian that describes the coherent part of the system evolution 

\begin{equation}
\begin{split}
H_{S}^{ef}&=  \Delta_{xL}\hat{\sigma}^+\hat{\sigma}^-+  \Delta_{cL}\hat{a}^{\dagger}a\\ &+\langle B\rangle \zeta_x+  \Delta^{\hat{\sigma}_11}_{W}\hat{\sigma}^+\hat{\sigma}^-\\ &+ \Delta_{ph}^{\hat{\sigma}^+\hat{a}}\hat{a}^\dagger\hat{\sigma}^-\hat{\sigma}^+\hat{a}
+  \Delta_{ph}^{\hat{\sigma}^-}\hat{\sigma}^-\hat{\sigma}^+
\\ &+  \Delta_{ph}^{\hat{a}^\dagger\hat{\sigma}^-}a\hat{\sigma}^+\hat{\sigma}^-\hat{a}^\dagger
+  \Delta_{ph}^{\hat{\sigma}^+}\hat{\sigma}^+\hat{\sigma}^-,
\end{split}
\end{equation}

and of the dissipative Lindbladian $\mathcal{L}_{ph}(\hat{\rho})$ and the coherent variational shift $D_{ph}(\hat{\rho})$. The former is defined according to 

\begin{equation}
\begin{split}
\mathcal{L}_{ph}(\hat{\rho})&=\frac{\Gamma_W^{\hat{\sigma}_{11}}}{2}L(\hat{\sigma}_{11})
+\mathcal{L}^{Intp}_{ph}+\frac{\Gamma_{ph}^{\hat{\sigma}^+\hat{a}}}{2}L(\hat{\sigma}^+\hat{a})\\
&+
\frac{\Gamma_{ph}^{\hat{\sigma}^-}}{2}L(\hat{\sigma}^-) +
\frac{\Gamma_{ph}^{\hat{a}^\dagger\hat{\sigma}^-}}{2}L(\hat{a}^\dagger\hat{\sigma}^-) + \frac{\Gamma_{ph}^{\hat{\sigma}^+}}{2}L(\hat{\sigma}^+) \hspace{1ex} ,
\end{split}
\end{equation}

where $\hat{\sigma}_{11} \equiv \hat{\sigma}^+ \hat{\sigma}^-$ and $L(\hat{D})=2\hat{D}\hat{\rho} \hat{D}^{\dagger}-\hat{D}^{\dagger}\hat{D}\hat{\rho}-\hat{\rho} \hat{D}^\dagger \hat{D}$. 

The term $\mathcal{L}^{Intp}_{ph}(\hat{\rho})$ describes the incoherent interpolation processes between the weak coupling approach \cite{Harsij}, and the polaronic theory \cite{roy2011phonon}. It explicitly reads

\begin{equation}
\begin{split}
&\mathcal{L}^{Intp}_{ph}(\hat{\rho})=\\
&\ \ \ \ \frac{\Gamma_{zy}^{\hat{\sigma}_{11} \hat{\sigma}^+}}{2} L_{ph}^{Intp}(\hat{\sigma}_{11}, \hat{\sigma}^+) +  \Gamma_{zy}^{\hat{\sigma}_{11}\hat{\sigma}^-} L_{ph}^{Intp}({\hat{\sigma}_{11}, \hat{\sigma}^-})\\
&+\frac{\Gamma_{zy}^{\hat{\sigma}_{11} (\hat{\sigma}^+\hat{a})}}{2} L_{ph}^{Intp}(\hat{\sigma}_{11}, \hat{\sigma}^+\hat{a})+	\frac{\Gamma_{zy}^{\hat{\sigma}_{11}(\hat{\sigma}^-\hat{a}^{\dagger})}}{2} L_{ph}^{Intp}({\hat{\sigma}_{11}, \hat{\sigma}^-\hat{a}^{\dagger}})\\	
&	+\frac{\Gamma_{yz}^{\hat{\sigma}^+ \hat{\sigma}_{11}}}{2}
L_{ph}^{Intp}(\hat{\sigma}^+, \hat{\sigma}_{11}) + \frac{\Gamma_{yz}^{\hat{\sigma}^-\hat{\sigma}_{11}}}{2}
L_{ph}^{Intp}(\hat{\sigma}^-,\hat{\sigma}_{11})\\
&+\frac{\Gamma_{yz}^{(\hat{\sigma}^+\hat{a}) \hat{\sigma}_{11}}}{2}
L_{ph}^{Intp}(\hat{\sigma}^+\hat{a}, \hat{\sigma}_{11}) +\frac{\Gamma_{yz}^{(\hat{\sigma}^-\hat{a}^{\dagger})\hat{\sigma}_{11}}}{2}
L_{ph}^{Intp}(\hat{\sigma}^-\hat{a}^{\dagger}. \hat{\sigma}_{11}) \hspace{1ex} .
\end{split}
\end{equation}

with $L_{ph}^{Intp}({A},{B})={A}{B}\hat{\rho}(t)-\hat{\rho}(t){B}^{\dagger}{A}^{\dagger}-{B}\hat{\rho}(t){A}+{A}^{\dagger}\hat{\rho}(t){B}^{\dagger}$.

As for the variational coherent shift (which is also originated from interpolation between the weak coupling and the polaronic models) \cite{Royeff}, it is given by

\begin{equation}
\begin{split}
&D_{ph}(\hat{\rho})=  \Delta_{zy}^{\hat{\sigma}_{11} \hat{\sigma}^+} \mathfrak{D}_{ph}^{Intp}(\hat{\sigma}_{11}, \hat{\sigma}^+) +  \Delta_{zy}^{\hat{\sigma}_{11}\hat{\sigma}^-} \mathfrak{D}_{ph}^{Intp}({\hat{\sigma}_{11}, \hat{\sigma}^-})\\
&+  \Delta_{zy}^{\hat{\sigma}_{11} \hat{\sigma}^+\hat{a}} \mathfrak{D}_{ph}^{Intp}(\hat{\sigma}_{11}, \hat{\sigma}^+\hat{a})+	  \Delta_{zy}^{\hat{\sigma}_{11}\hat{\sigma}^-\hat{a}^{\dagger}} \mathfrak{D}_{ph}^{Intp}({\hat{\sigma}_{11}, \hat{\sigma}^-\hat{a}^{\dagger}})\\	
&	+  \Delta_{yz}^{\hat{\sigma}^+ \hat{\sigma}_{11}}
\mathfrak{D}_{ph}^{Intp}(\hat{\sigma}^+, \hat{\sigma}_{11}) +   \Delta_{yz}^{\hat{\sigma}^-\hat{\sigma}_{11}}
\mathfrak{D}_{ph}^{Intp}(\hat{\sigma}^-,\hat{\sigma}_{11})\\
&+  \Delta_{yz}^{\hat{\sigma}^+\hat{a} \hat{\sigma}_{11}}
\mathfrak{D}_{ph}^{Intp}(\hat{\sigma}^+\hat{a}, \hat{\sigma}_{11}) +  \Delta_{yz}^{\hat{\sigma}^-\hat{a}^{\dagger}\hat{\sigma}_{11}}
\mathfrak{D}_{ph}^{Intp}(\hat{\sigma}^-\hat{a}^{\dagger}, \hat{\sigma}_{11}) \hspace{1ex},\\
\end{split}
\end{equation}

where $\mathfrak{D}_{ph}^{Intp}({A},{B})={A}{B}\hat{\rho}(t)+\hat{\rho}(t){B}^{\dagger}{A}^{\dagger}-{B}\hat{\rho}(t){A}-{A}^{\dagger}\hat{\rho}(t){B}^{\dagger}$.

By comparing equation (\ref{simplifiedME}) with (\ref{VME}), and dropping highly oscillatory terms, we obtained the phonon mediated transition probabilities and the variational shifts. The thermal dissipative rates are found to be of three types: Weak coupling-like rates \cite{Harsij,WeakIran}



\begin{align}
&	\Gamma_{W}^{\hat{\sigma}_{11}}=2    \Re\left[\int_0^\infty d\tau Czz(\tau)\right] \hspace{1ex}, 
\end{align}

polaronic-like rates \cite{Royeff} 

\begin{align}
&\Gamma_{ph}^{\hat{\sigma}^+\hat{a}/\hat{a}^\dagger\hat{\sigma}^-}=2   g^2\Re\left[\int_0^\infty d\tau \langle B\rangle ^2 e^{\pm \Delta_{cx}\tau}\left(e^{\phi(\tau)}-1\right)\right]	,\\
&\Gamma_{ph}^{\hat{\sigma}^+/\hat{\sigma}^-}=2  \eta_x^2\Re\left[\int_0^\infty d\tau \langle B\rangle ^2 e^{\mp \Delta_{xL}\tau}\left(e^{\phi(\tau)}-1\right)\right] \hspace{1ex},	
\end{align} 

and interpolated rates 
 
\begin{align}
&\Gamma^{\hat{\sigma}_{11}\hat{\sigma}^{\pm}}_{zy}= \mp2    \eta_x\Im\left[\int_{0}^{\infty}d\tau  C_{zy}(\tau)e^{\mp\Delta_{XL}\tau}\right],\\
&\Gamma^{\hat{\sigma}_{11} (\hat{\sigma}^+\hat{a}/\hat{\sigma}^-\hat{a}^{\dagger})}_{zy}=\mp 2    g  \Im\left[\int_0^{\infty}d\tau C_{zy}(\tau)e^{\pm\Delta_{CX}\tau}\right],\\
&\Gamma^{\hat{\sigma}^{\pm}\hat{\sigma}_{11}}_{yz}= \mp2    \eta_x\Im\left[\int_{0}^{\infty}d\tau  C_{yz}(\tau)\right],\\
&\Gamma^{ (\hat{\sigma}^+\hat{a}/\hat{\sigma}^-\hat{a}^{\dagger}) \hat{\sigma}_{11}}_{yz}=\mp2    g  \Im\left[\int_0^{\infty}d\tau C_{yz}(\tau)\right] \hspace{1ex}.
\end{align}

Meanwhile, the energy shift components are identified as
	
\begin{align}
&	\Delta_{W}^{\hat{\sigma}_{11}}=\Im\left[\int_0^\infty d\tau  Czz(\tau)\right]\\	 
&\Delta_{ph}^{\hat{\sigma}^+\hat{a}/\hat{a}^\dagger\hat{\sigma}^-}= g^2\Im\left[\int_0^\infty d\tau \langle B\rangle ^2 e^{\pm \Delta_{cx}\tau}\left(e^{\phi(\tau)}-1\right)\right]	,\\
&\Delta_{ph}^{\hat{\sigma}^+/\hat{\sigma}^-}= \eta_x^2\Im\left[\int_0^\infty d\tau \langle B\rangle ^2e^{\mp \Delta_{xL}\tau}\left(e^{\phi(\tau)}-1\right)\right],	\\
&\Delta^{\hat{\sigma}_{11}\hat{\sigma}^{\pm}}_{zy}= \pm \eta_x\Re\left[\int_{0}^{\infty}d\tau  C_{zy}(\tau)e^{\mp\Delta_{XL}\tau}\right],\\
&\Delta^{\hat{\sigma}_{11}\ \hat{\sigma}^+\hat{a}/\hat{\sigma}^-\hat{a}^{\dagger}}_{zy}=\pm g  \Re\left[\int_0^{\infty}d\tau C_{zy}(\tau)e^{\pm\Delta_{CX}\tau}\right],\\
&\Delta^{\hat{\sigma}^{\pm}\hat{\sigma}_{11}}_{yz}= \pm \eta_x\Re\left[\int_{0}^{\infty}d\tau  C_{yz}(\tau)\right],\\
&\Delta^{\hat{\sigma}^+\hat{a}/\hat{\sigma}^-\hat{a}^{\dagger}\ \hat{\sigma}_{11}}_{yz}=\pm g  \Re\left[\int_0^{\infty}d\tau C_{yz}(\tau)\right].
\end{align}


	\section{V. NUMERICAL RESULTS}
	
	As a representative study case, we will focus on the resonance fluorescence of a InAs/GaAs quantum dot coupled to a high quality optical resonator, under resonant continuous wave excitation \cite{HYR1,HYR2}. It is known that in most III-V semiconductor materials, the main source of dephasing is the carrier-acoustic phonon interaction via deformation potential \cite{krummheuer2002theory,Gauger}. Thus, the spectral density  $J_{ph}(\omega)=\alpha\omega^3e^{-\omega^2/2\omega^2_b}$, is adopted for the simulations. $\alpha$ captures the strength of the exciton-phonon coupling and $\omega _b$ provides a natural high-frequency cutoff, which is proportional to the inverse of the carrier localization length in the QD \cite{nazir2016modelling}.
	
	To simulate the fluorescence spectrum, we compute 
		
		\begin{equation}
		\begin{split}
		S_c(\omega)\propto \lim_{t\rightarrow\infty}\Re\bigg[&\int_{0}^{\infty}d\tau\bigg[\langle a(t+\tau)\hat{a}^{\dagger}(t)\rangle\\ 
		& -\langle a(t+\tau)\rangle  \langle \hat{a}^{\dagger}(t)\rangle \bigg]e^{i(\omega_L-\omega)\tau}\bigg],
		\end{split}      
		\end{equation}
	
	where the correlation functions are obtained by the quantum regression formula \cite{carmichael2000statistical}. To numerically solve the master equation within the different levels of approximation compared here (weak coupling, polaronic and variational), we employ a quantum optics toolbox developed in \textit{MATLAB} by Tan S. M. \cite{tan1999computational}. The pumping rate is assumed stable, i.e. $\eta_x$ is taken independent of time, and the emitter is considered in the base state as initial condition \cite{perea2004dynamics}. 
	
	To make our results comparable with Mollow triplet experiments on semiconductor micropillars by S. M. Ulrich {\it et al.} \cite{ulrich2011dephasing}, we consider a mode-cavity detuning  $\omega_c-\omega_x=-0.2$ meV, and a radioactive decay rate $ \gamma=3$ $\mu$eV. Those values are also similar to the ones used in experiments by F. Hargart {\it et al.} in reference \cite{Hargart} and by H. Kim {\it et al.} in reference \cite{PRLKim}. 
	
	As for phonon parameters, typical values for InAs/GaAs QDs are used (cutoff frequency $\omega_b=0.9$ meV y $\alpha_p=0.03$ ps$^{2}$) \cite{Hargart,ramsay2010damping,ramsay2010phonon}.

\begin{figure*}
\includegraphics[width=0.33\linewidth]{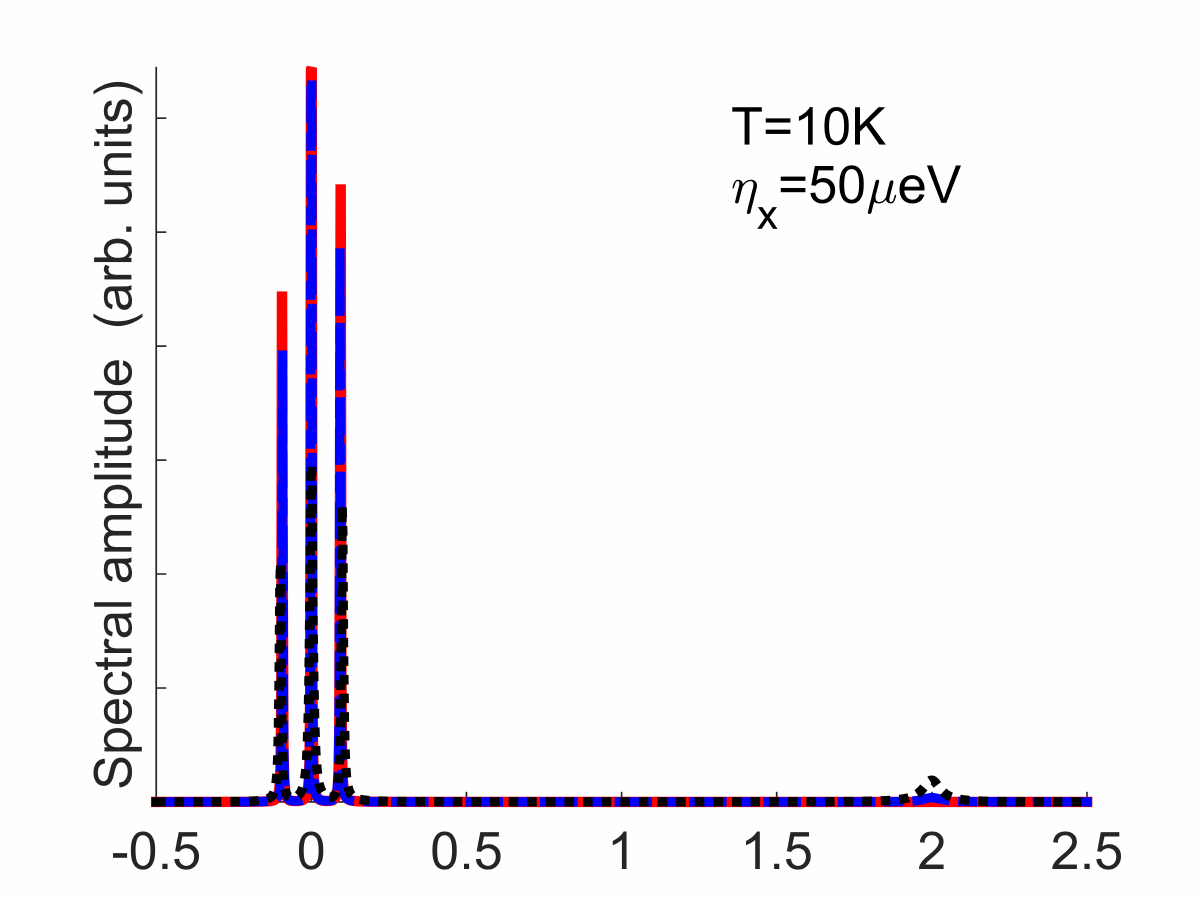}\includegraphics[width=0.33\linewidth]{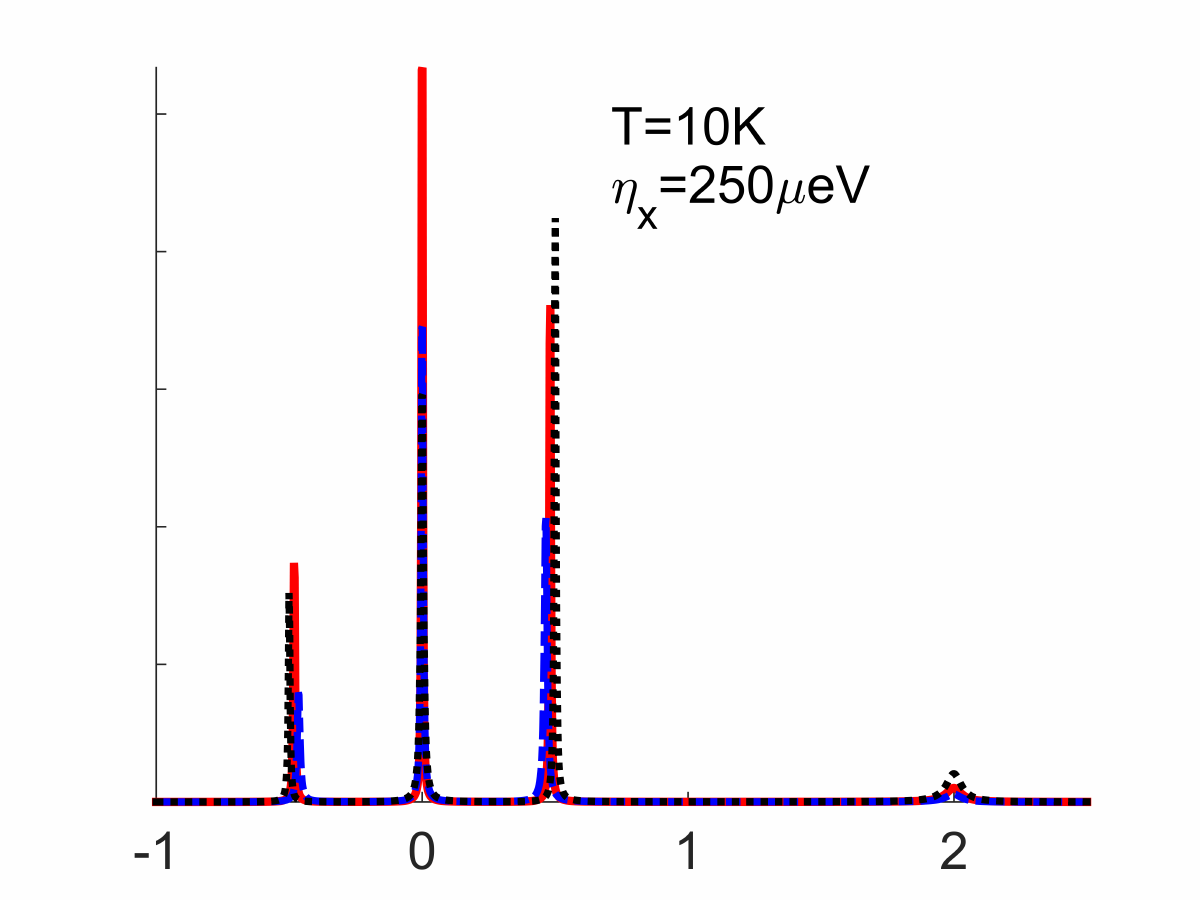}\includegraphics[width=0.33\linewidth]{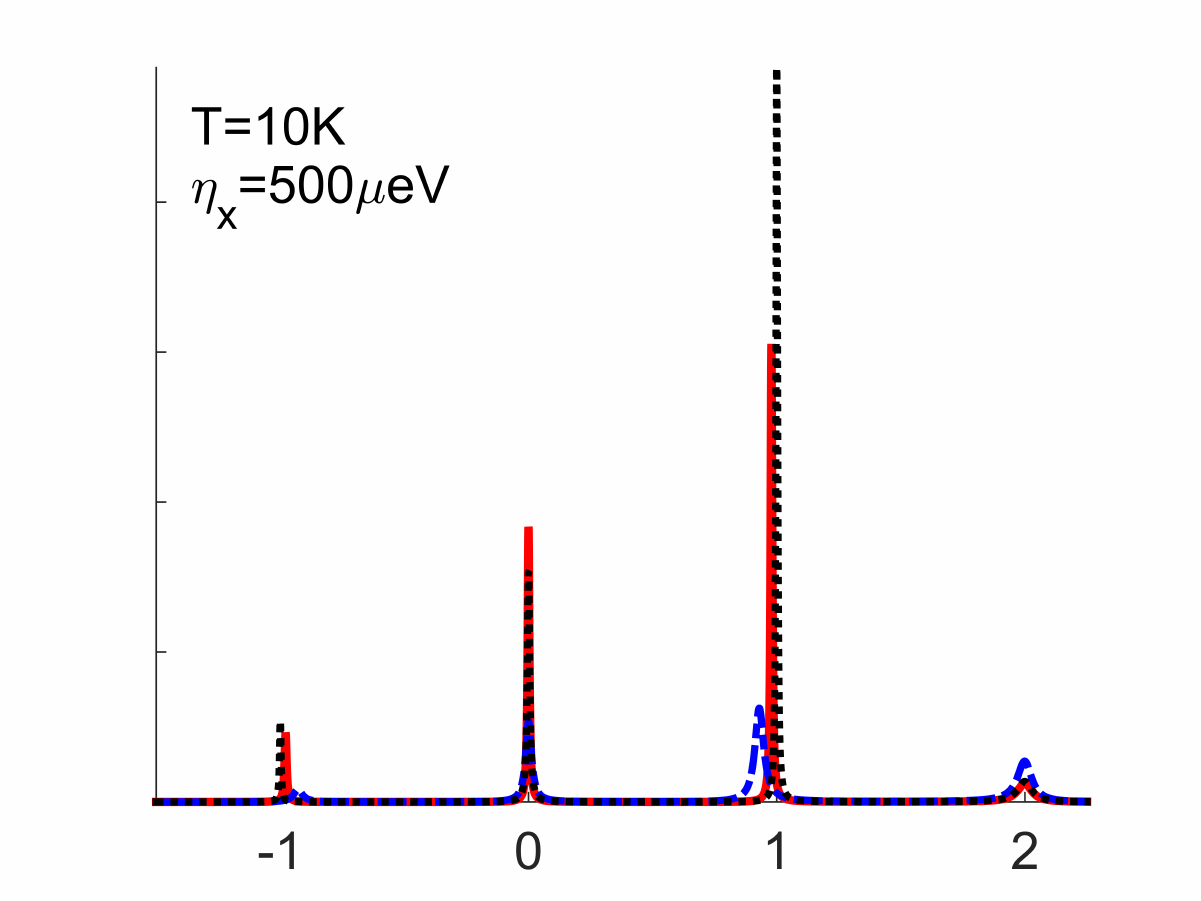}\\
\includegraphics[width=0.33\linewidth]{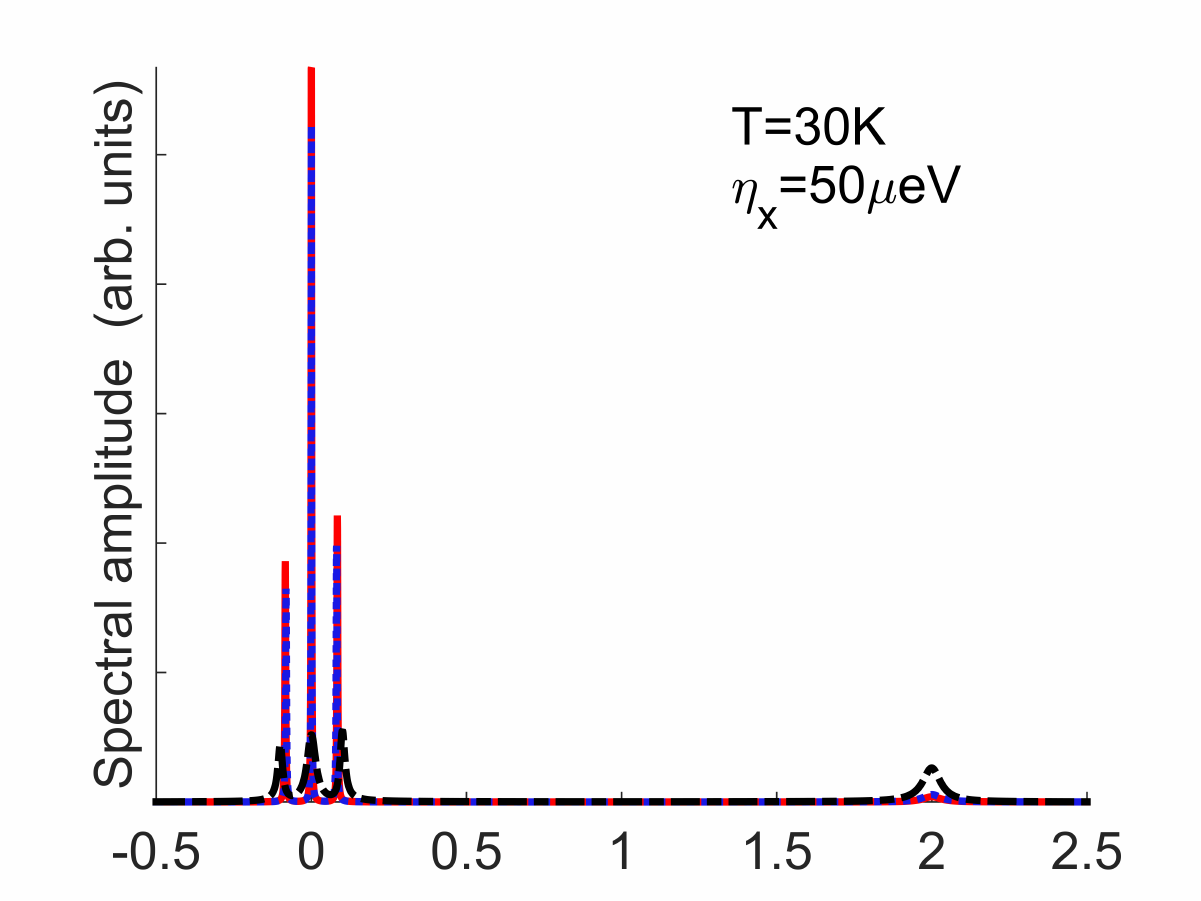}\includegraphics[width=0.33\linewidth]{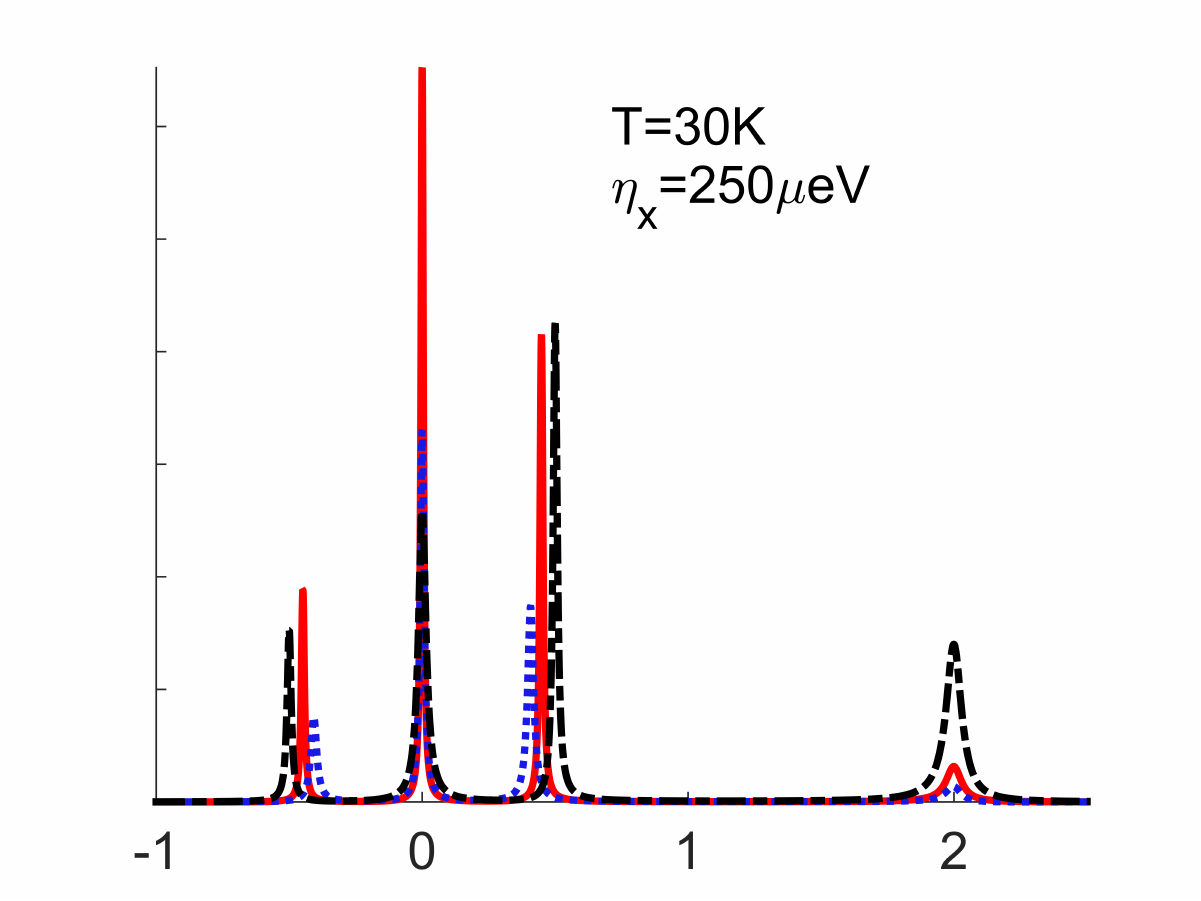}\includegraphics[width=0.33\linewidth]{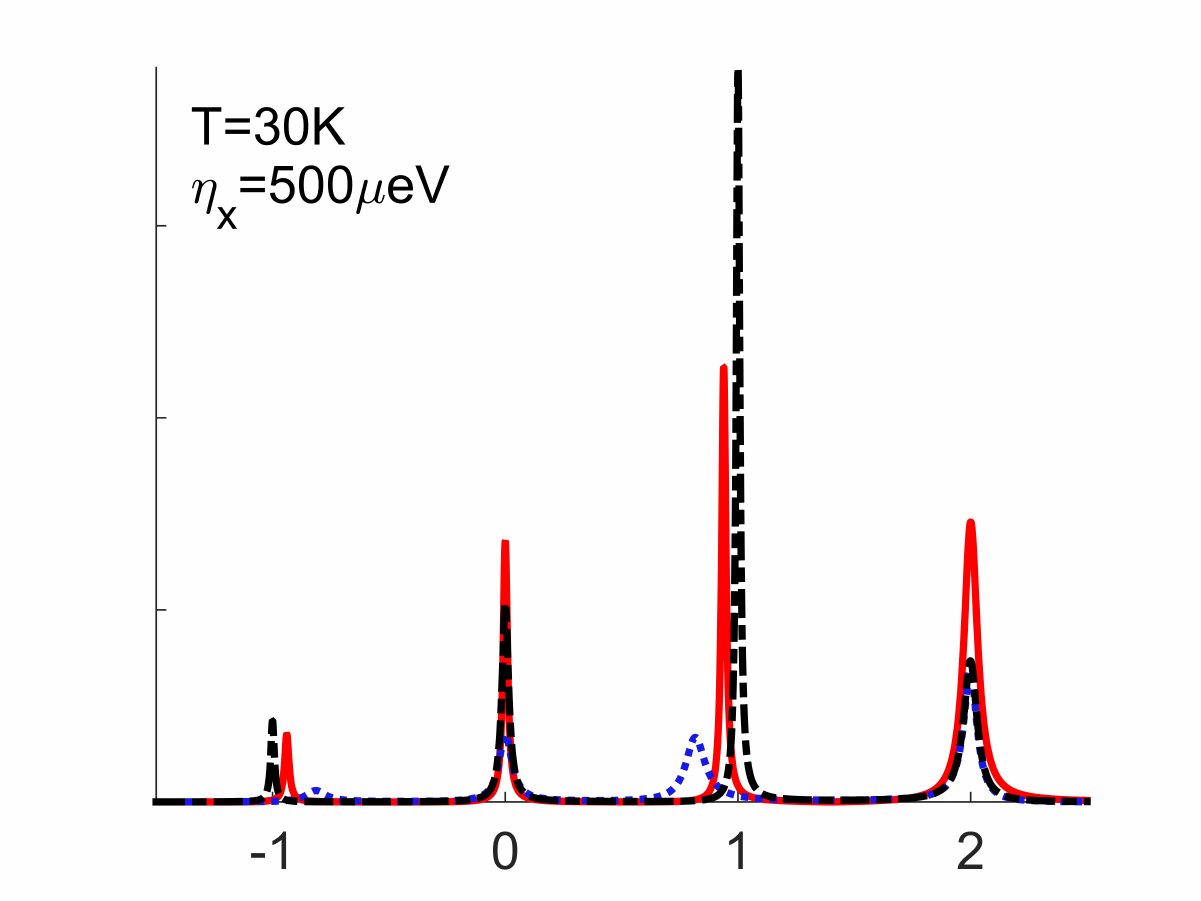}\\
\includegraphics[width=0.33\linewidth]{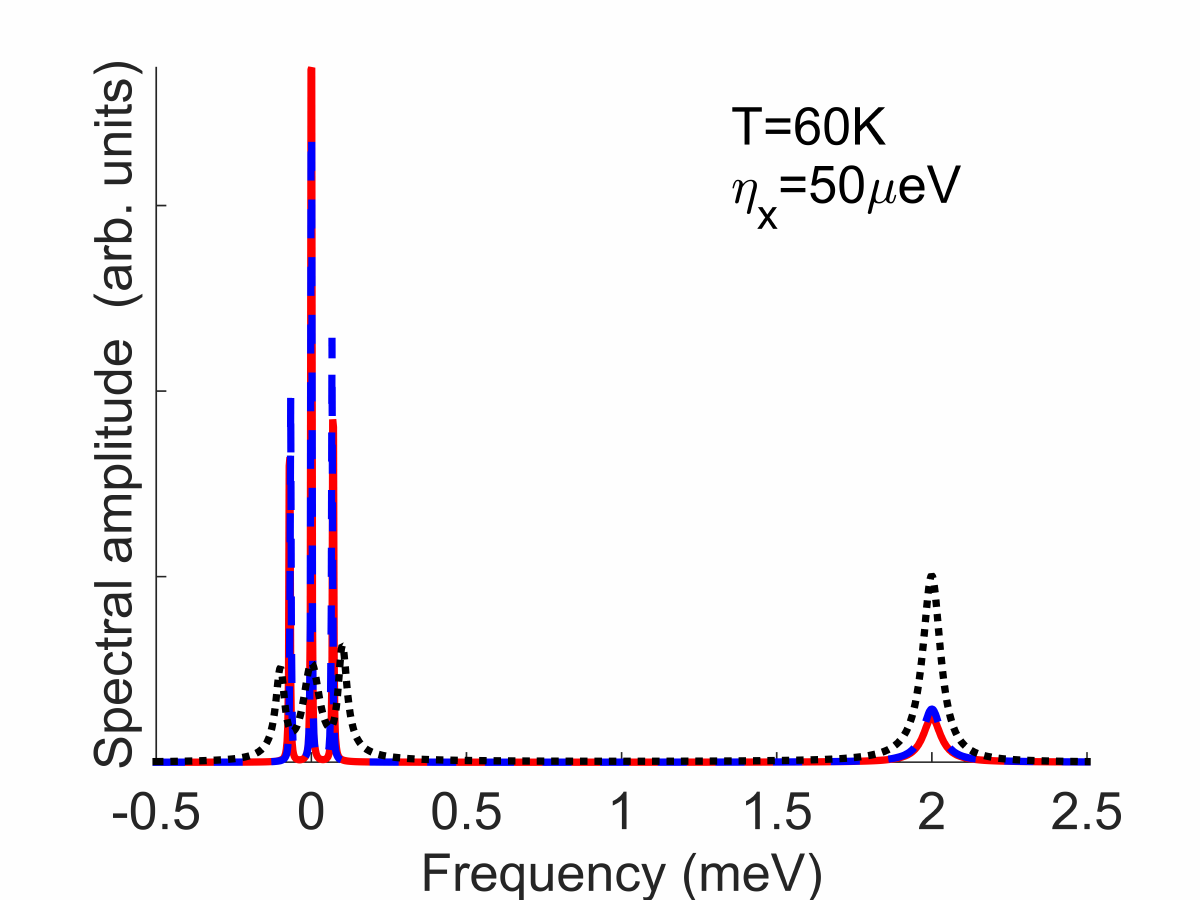}\includegraphics[width=0.33\linewidth]{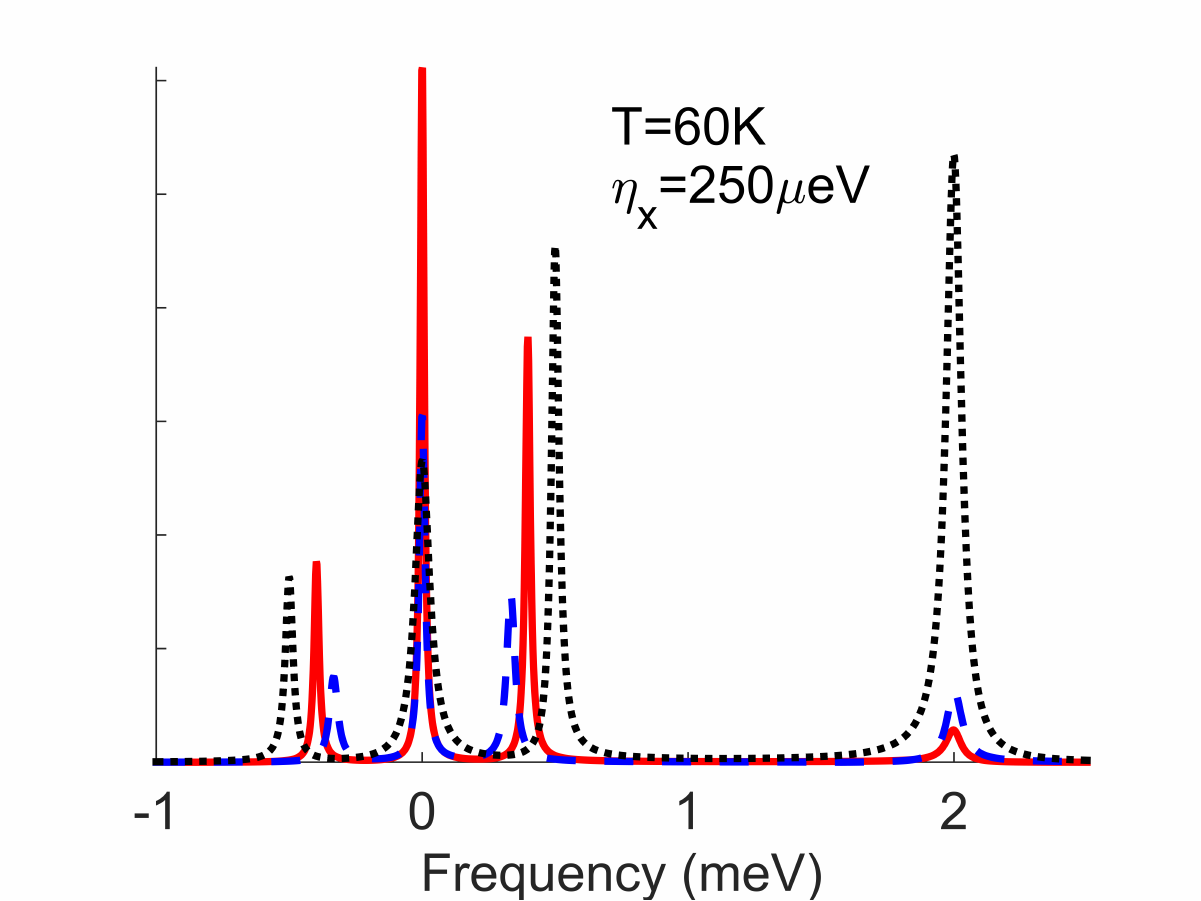}\includegraphics[width=0.33\linewidth]{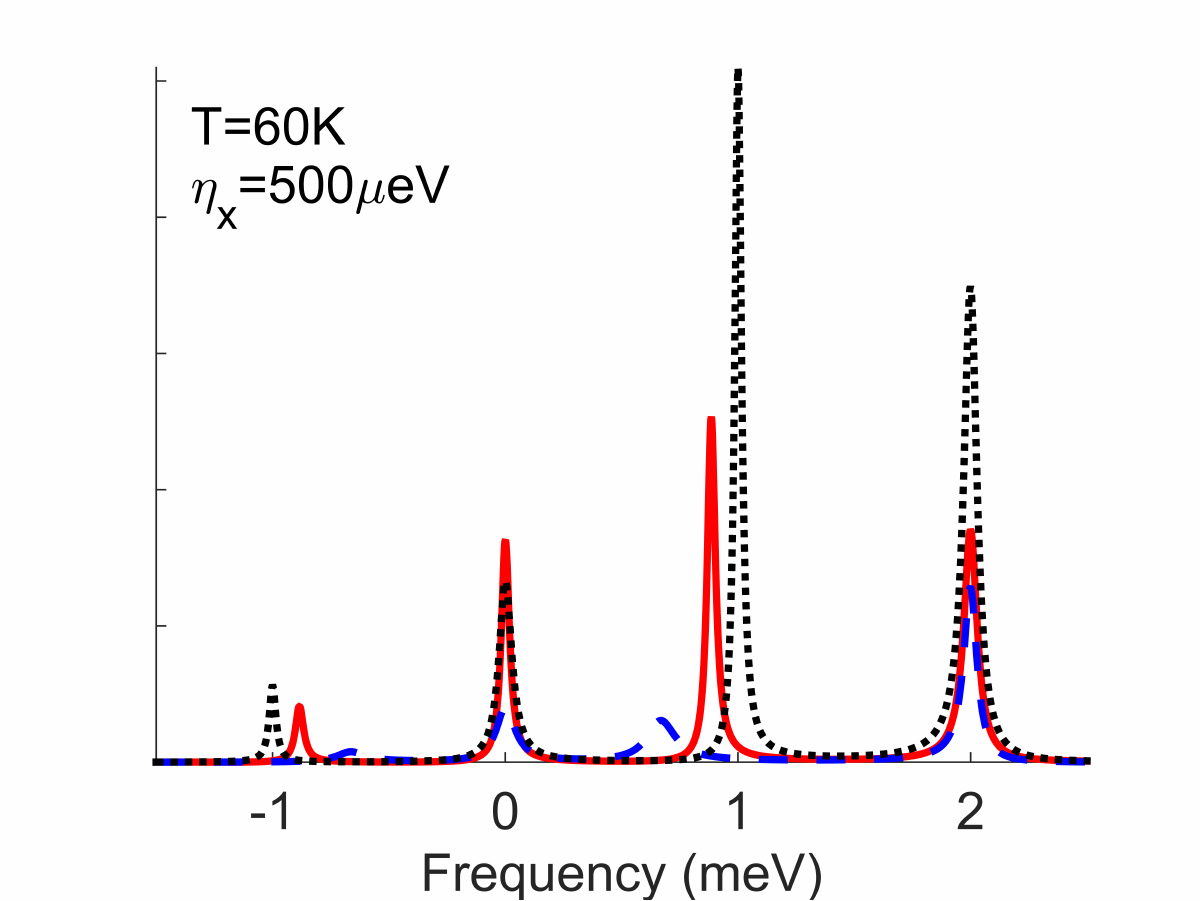}
\caption{Cavity-emitted fluorescence spectra of a semiconductor QD-cavity system driven via on-resonance exciton pumping ($\omega_L = \omega_x$, with $\Delta_{cx} = 2$ meV) for various values of the exciton pump $\eta_x$ and phonon-bath temperature $T$. Black line: spectra obtained from a weak coupling master equation, blue line: from a polaronic master equation, and red: from the variational master equation developed in this work. In all plot panels, the frequency is taken respect to the QD emission and $g=26.7$ $\mu$eV is used.} 
\label{esp}
\end{figure*}

Figure \ref{esp} shows emission spectra from the cavity under various pumping rates and temperatures, obtained within the three considered master equation approaches.     

One can see how the weak coupling model differs greatly from the polar and variational theories as the system temperature increases, because of overestimation of the phonon dissipative effects. Concurrently, as long as the pumping rate remains moderate (e.g. $\eta_x=50$ $\mu$eV), the polar and variational approaches predict similar behaviors. In this regime, the polaron model has been successfully fitted to resonance fluorescence measurements \cite{hefeitemperature}. However, contrasts between those two later master equations are revealed when the pumping rate is strengthened. 

At median laser power (e.g. $\eta_x=250$ $\mu$eV), the variational theory exhibits intermediate results between the weak and the polaronic models, which is particularly observable at the Mollow triplet side peaks. 

Under high excitation conditions (e.g. $\eta_x=500$ $\mu$eV), the polaronic and variational approaches differ significantly in the predicted renormalization of the Rabi frequency and the emission intensity of all the peaks, specially the right one in the triplet, evidencing how in this regime the polaronic approach also misjudges by excess the phonon associated decoherence.  

Such a breakdown of the polaronic approach for high pumping rates becomes larger as the temperature increases. Surprisingly for strong pumping, as compared with the variational results, predictions from the weak coupling model differ less than those from the polaronic model.

In order to check consistency of our results with real-time path integral calculations, we compare the Rabi frequency renormalization in the bottom-right panel of figure 3, to those reported in figure 3b) of reference \cite{PRB2012} and figure 5 of reference \cite{PRB2014}. There, a renormalization of $\sim 10$\% is reported for bare Rabi frequencies at the order of 1 meV, in agreement with our simulations from the variational model, while such a renormalization obtained within the polaron approach reaches $\sim 35$\%, elucidating  overvaluation of the thermal effects.

It is worth mentioning that in despite of discrepancies regarding its magnitude, all three models account for the phonon assisted cavity feeding phenomenon.  

\section{VI. Summary and conclusions}
	
In this work, we derived an optimized master equation for a quantum photon emitter simultaneously coupled to a phonon bath and to an optical resonator, inspired on the polaronic transformation but with phonon displacements variationally determined by a mode-dependent approach. Thus, a theory flexible enough to encompass the weak and polaronic coupling methods, but applicable on a larger range of experimental conditions, was obtained.

We applied the developed theory in the simulation of the resonance fluorescence emission from a single quantum dot embedded into a high quality microcavity, for different temperature and excitation values. Such spectra were also calculated within the weak coupling and conventional polaronic theories, so that pertinent comparison could be carried out among the three considered models.
 
The numerical results showed that in comparison to the more rigorous variational approach, the weak coupling and polaronic theories, correspondingly overestimate the phonon dissipative effects as the temperature and the excitation power increase.   
  
In conclusion, the variational master equation obtained here, is expected to provide a valuable tool to simulate and explain experiments on solid-state emitters interacting with phonon reservoirs and QED cavities, carried out under light-matter coupling, pumping rate and temperature values, lying in a much wider range than those spanned by previously available master equation approaches. This is of significance given the increasing excitation intensities and emitter-cavity mode couplings achieved in state of art quantum optical experiments. 
 
	
\section{Acknowledgments}
The authors acknowledge the Research Division of Universidad Pedag\'ogica y Tecnol\'ogica de Colombia for financial support.

	
	
	

	\bibliography{References}

	
\end{document}